\documentclass[11pt,a4paper]{article}
\usepackage{jheppub_kim}

\usepackage{pdflscape}
\usepackage{amsmath}
\usepackage{amssymb}
\usepackage{dcolumn}
\usepackage{bm}
\usepackage{color}
\usepackage{epsfig}
\usepackage{amsfonts}
\usepackage{graphicx}
\usepackage{subfigure}
\usepackage{dcolumn}

\begin{document}

\title{Higher order corrected thermodynamics and statistics of Kerr-Newman-G\"{o}del black hole}
\author[a]{B. Pourhassan,}
\author[a]{K. Kokabi,}
\author[a]{Z. Sabery}

\affiliation[a]{School of Physics, Damghan University, Damghan, 3671641167, Iran}

\emailAdd{b.pourhassan@du.ac.ir}
 \emailAdd{kokabi@du.ac.ir}
\emailAdd{z.sabery@std.du.ac.ir}

\abstract{In this paper, we consider the rotating charged G\"{o}del black hole  and study the effect of the higher order corrections of the entropy on the thermodynamics and statistics quantities of the Kerr-Newman-G\"{o}del black hole. The leading order correction is logarithmic while higher-order terms are proportional to the inverse of the area of the black hole. We obtain modified thermodynamics and statistics and find that correction terms are important in the stability of the black hole.}


\maketitle

\section{Introduction}
Many studies have been done on the black hole physics. The black attribute in the black hole is because it captures all the light that passes through its event horizon, from which the black hole behaves like a black body in thermodynamics \cite{01}, hence they have
blackbody radiation, which yields to the semi-classical methods of Hawking radiation as a tunneling process \cite{PLB1}.\\
According to Hawking's theory, the entropy of a black hole is proportional to the area of the black hole, and it is proportional to the power of the radius of the event horizon. In that case, black hole thermodynamics is one of the important fields of research in theoretical physics \cite{CQG1, CQG2, CQG3, CQG4, CQG5}. Small statistical disturbances around the equilibrium point produce correction terms in the black hole entropy \cite{03}. The lowest entropy corrections are logarithmic \cite{04} and higher order corrections are proportional to the powers of the inverse of the black hole area \cite{05}.\\
Investigating the thermodynamics of black holes in the presence of the correction terms, gives us important information about the black hole. Indeed study about the modified entropy increases our quantum knowledge and may help to realize the theory of quantum gravity. These correction terms in the black hole entropy for a large-sized cases are trivial and can be ignored, while they are important for the small black holes, and give important results \cite{06}.\\
Several types of black objects have been investigated by the presence of these correction terms of the black hole entropy \cite{07, 08, 09, 010, 011}. Leading order thermal fluctuations in a charged AdS black hole considered by the Ref. \cite{12}, which for the first time general coefficient for the logarithmic correction of the form $\ln{S_{0}T^{2}}$ has been introduced, where $S_{0}$ is the original entropy of the black hole with Hawking temperature $T$. Leading-order quantum correction to the semi-classical black hole entropy for the first time calculated by the Ref. \cite{04} as $\ln S_{0}$. In that case, it has been suggesting that the coefficient of the correction may be universal \cite{7}. The mentioned form of the logarithmic corrections to the rotating extremal black hole entropy in four and five dimensions \cite{8} as well as Schwarzschild and other
non-extremal black holes in different dimensions \cite{9} already studied. This kind of logarithmic correction also used to study simple regular black hole solution satisfying the weak energy condition \cite{10}. It is  calculated from three-dimensional black holes with soft hairy boundary conditions \cite{new2}. The same logarithmic corrections to the entropy of black hole obtained from Kerr/CFT correspondence \cite{11}.\\
P-V criticality of first-order entropy corrected AdS black holes in massive gravity recently studied by the Ref. \cite{PRD22}. Such leading order correction \cite{EPJC22} also considered in a Hyperscaling violation background \cite{EPJC33, EPJC44, EPJC55}.\\
The effect of the logarithmic term on the entropy functional formalism already investigated by Ref. \cite{15} and found that the leading correction to the micro-canonical entropy may be used to recover modified theories of gravity such as $f(R)$ theory of gravity \cite{16, 17}. The effects of leading order thermal
fluctuations recently studied for the Reissner-Nordstr\"{o}m-AdS black hole \cite{21} which show that critical exponents are the same as critical exponents without thermal fluctuations.\\
Logarithmic corrected entropy is also important from AdS/CFT point of view \cite{23}. In the Ref. \cite{23} it has been found that the lower bound violation of shear viscosity to entropy ratio \cite{24,25,26,27,28,29} due to logarithmic correction may be violated.\\
One of the most interesting kinds of black holes is G\"{o}del black hole \cite{012, PLB2}. In this paper, we consider the G\"{o}del black hole with electric charge and rotation \cite{013} and examine the thermodynamic and statistical parameters of a black hole with higher order terms in the black hole entropy. In that case, thermodynamics and statistics of ordinary (non-rotating, uncharged) G\"{o}del black hole have been studied by the Ref. \cite{IJTP1}. Also, Ref. \cite{IJTP2} have been studied another case of logarithmic correction of the entropy on the thermodynamics and statistics of Kerr-G\"{o}del black hole. Now, we would like to obtain corrected thermodynamics and statistics quantities of Kerr-Newman-G\"{o}del black hole in the presence of higher order corrections.\\
This paper is organized as follows. In the next section, we review some important properties of Kerr-Newman-G\"{o}del black hole. In section 3 we write corrected entropy and study corrected thermodynamics of Kerr-Newman-G\"{o}del black hole. We obtain behavior of some important thermodynamics quantities like internal and Helmholtz free energies, enthalpy and Gibbs free energy. In section 4 we obtain the relation between pressure and volume which is useful to study critical points. In section 5 we analyze specific heat to investigate thermodynamics stability of the black hole. In section 6 we have the statistical study for the Kerr-Newman-G\"{o}del black hole. Finally, in section 7 we give the conclusion.

\section{Kerr-Newman-G\"{o}del black hole}
Kerr-Newman-G\"{o}del black hole in 5-dimensional space-time is described by the following metric \cite{014},
\begin{eqnarray}\label{1}
{d}{s}^2=&-&{f}{(r)}[{d}{t}+{\frac{{h}{(r)}}{{f}{(r)}}}(d{\phi}+{\cos}{\theta}{d\psi})]^2\nonumber\\
&+&{1 \over 4}{r^2}({d\theta^2}+{\sin^2}{\theta}{d\psi^2})+{{d}{r^2} \over {v}{(r)}}+{{r^2}{v}{(r)} \over {4}{f}{(r)}}{({d\phi}+{\cos}{\theta}{d}{\psi})^2},
\end{eqnarray}
where the Euler angle $\theta$, $\psi$ and $\phi$ run over the ranges $0$ to $\pi$, $0$ to $2\pi$, and $0$ to $4\pi$, respectively. Also \cite{1101},
\begin{equation}\label{2}
{f}{(r)}={1}-{\frac{2\mu}{r^2}}+{\frac{q^2}{r^4}},
\end{equation}

\begin{equation}\label{3}
{h}{(r)}={j}{r^2}+{3}{j}{q}+{{(2{\mu}-{q})}{a} \over {2}{r^2}}-{{q^2}{a} \over {r^4}}
\end{equation}
and
\begin{eqnarray}\label{4}
{v}{(r)}&=&{1}-{{2}{\mu} \over {r^2}}+{{8}{j}({\mu}+{q})[{a}+{2}{j}({\mu}+{2}{q})] \over {r^2}}\nonumber\\
&+&{{2}({\mu}-{q}){a^2} \over {r^4}}+{{q^2}[{1}-{16}{j}{a}-{8}{j^2}({\mu}+{3}{q})] \over {r^4}},
\end{eqnarray}
where, $mu$ denotes black hole mass, $q$ is electrical charge and $j$ represent the G\"{o}del parameter, which is responsible for the rotation of the G\"{o}del universe \cite{G}. The angular velocities  the G\"{o}del black hole are given by \cite{1101},
\begin{equation}\label{5}
{\Omega}{(r)}={\Omega}_{\phi}={{h}{(r)} \over {U}{(r)}},
\end{equation}
where,
\begin{eqnarray}\label{6}
{U}{(r)}&=&{{{r^2}{v}{(r)}-{4}{h^2}{(r)}} \over {4}{f}{(r)}}\nonumber\\
&=&-{j^2}{r^2}({r^2}+{2}{\mu}+{6}{q})+{3}{j}{q}{a}+{({\mu}-{q}){a^2} \over {2}{r^2}}-{{q^2}{a^2} \over {4}{r^2}}+{{r^2} \over {4}}.
\end{eqnarray}
The outside event horizon ${r}_{+}$ is determined by equation ${v}({r}_{+})=0$ which yields to the following relation,
\begin{eqnarray}\label{7}
{r}_{+}^{2}={\mu}-{4}{j}({\mu}+{q}){a}-{8}{j^2}({\mu}+{q})({\mu}+{2}{q})+\sqrt{\delta}
\end{eqnarray}
where
\begin{eqnarray}\label{7-1}
{\delta}=\left({\mu}-{q}-{8}{j^2}{({\mu}+{q})^2}\right)\left({\mu}+{q}-{2}{a^2}-{8}{j}({\mu}+{2}{q})a-{8}{j^2}({\mu}+{2}{q})^2\right).
\end{eqnarray}
We will use above solutions to investigate the modified thermodynamics due to the higher order corrections of the black hole entropy.

\section{Black hole corrected thermodynamics}
We consider the black hole as a thermodynamic system. It has already been proved that the thermodynamic quantities of the G\"{o}del black hole apply to the first law of thermodynamics \cite{014}. The Hawking temperature of the Kerr-Newman-G\"{o}del black hole is given by the following formula \cite{1101},
\begin{equation}\label{8}
{T}={r}_{+}{v^{\prime}}({r}_{+}) \over {8}{\pi}{\sqrt{U({r}_{+})}},
\end{equation}
where $v$ is given by the equation (\ref{4}) and "prime" denote the derivative with respect to the black hole radius. Also its original (un-corrected) entropy is given by,
\begin{equation}\label{9}
{S}_{0}={\pi}^2{r}_{+}^2{\sqrt{U({r}_{+})}}.
\end{equation}
Knowing entropy in terms of the microstates $\Omega(E)$ as ${S}={\ln\Omega}(E)$, single particle energy spectrum for each quantum number given by,
\begin{equation}\label{10}
{E}_{n}={f}(n)
\end{equation}
while quantum density of state given by,
\begin{equation}\label{11}
{\rho}(E)={\Sigma_{n}}\Omega(E_{n})\delta(E-E_{n}),
\end{equation}
where delta function defined as,
\begin{equation}\label{12}
\delta(E-E_{n})=\delta(E-f(n))=\delta(n-F(E))|F\prime(E)|,
\end{equation}
and quantum density of state given by,
\begin{equation}\label{13}
{\rho}(E)={\Omega}({E})|F^{\prime}({E})|{\sum}_{n=0}^{n=\infty}{\delta}({n}-{F}({E})).
\end{equation}
One can deduce for each thermodynamic system that corrected $S(E)$ in the presence of thermal impairments would be calculated as follow,
\begin{eqnarray}\label{14}
{S}({E})&=&{S}_{0}({E})-{\ln}({F}({E}))-{1 \over 2}{\ln}({\alpha}_{2}^{2})+{\sum}_{n=2}^{n=\infty}{{{\alpha}_{2n}}({-1})^{n} \over ({2n}){!}{!}{{\alpha}_{2}}^{2}{n}}\nonumber\\
&+&{{1} \over {2}{!}}{\Sigma}_{n=3}^{\infty}{{\Sigma}_{m=3}^{\infty}}{{{{\alpha}_{n}}{{\alpha}_{m}}{({-1})^{k}}({2k}-{1}){!}{!} \over {n}{!}{m}{!}{{\alpha}_{2}}^{2}{k}}}+\mathcal{O}{(\frac{\alpha_{4}^{3}}{{\alpha}_{2}^{12}})}.
\end{eqnarray}
As subleading order terms have dependence on $m$ and $n$ i.e. order unity one can write corrected entropy as follow,
\begin{equation}\label{15}
{S}={S}_{0}-{{\alpha} \over {2}}\ln({S}_{0}{T^2})+{{\gamma} \over {S}_{0}}+\cdots,
\end{equation}
which is in a good agrement with what predicted in \cite{03}.
In this regard, the higher powers of ${1} \over {S}_{0}$ have been discarded. $\alpha$ and $\gamma$ are constants that have different values for different systems and  determined using system conditions. We examine the thermodynamics of the rotating charged G\"{o}del black hole with the entropy and Hawking temperature.\\

\begin{figure}[th]
\begin{center}
\includegraphics[scale=.5]{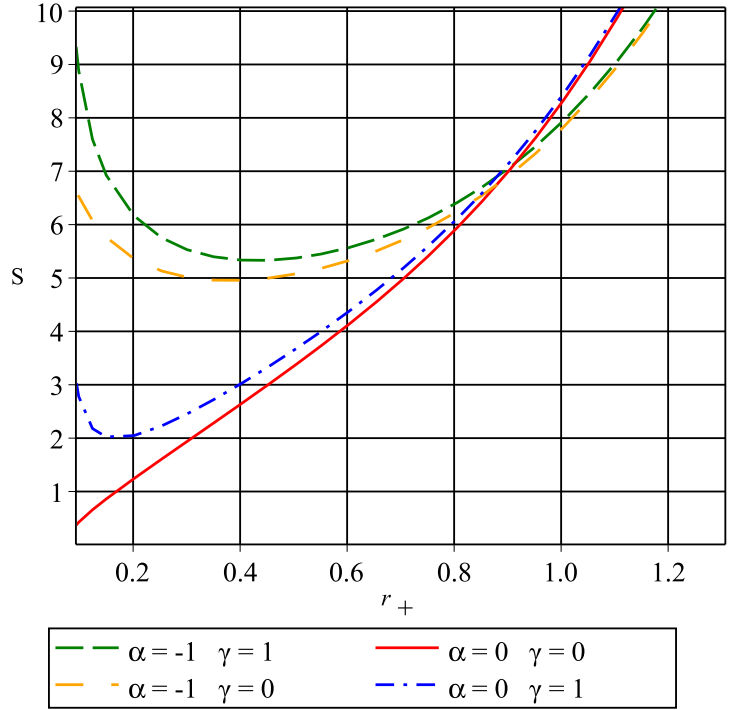}
\caption{Entropy according to the radius of the event horizon for values $\mu=1$, $q=0.1$, $a=1$, $j=0.0649$.}
\label{fig1}
\end{center}
\end{figure}

We give graphical analysis of the entropy for the several values of parameters (see Fig. \ref{fig1}). In the radius of a small horizon, in the absence of correction sentences, the graph is almost linear. By entering the logarithmic correction, the entropy changes and increases and in the presence of the second-order correction, the entropy shows a different behavior. Finally, in the presence of both corrective sentences, the entropy of the system is completely different, and the entropy of the system is greatly increased.

\subsection{Internal energy}
In the canonical ensemble, the relationship between entropy and the partition function is as follow,
\begin{equation}\label{16}
{S}(T)={k}_{B}{\ln}{Z}+{k}_{B}{T}{{\partial}{\ln Z} \over {\partial T}},
\end{equation}
which is yields to the following relation,
\begin{equation}\label{17}
{\ln Z}={{1} \over {T}}{\int}{{S}({T}) \over {k}{B}}{d}{T}
\end{equation}
The internal energy of the system is obtained using partition function as follow,
\begin{equation}\label{18}
{E}={k}_{B}{T}^{2}{{d}{\ln Z} \over {d}{T}}
\end{equation}

\begin{figure}[th]
\begin{center}
\includegraphics[scale=.3]{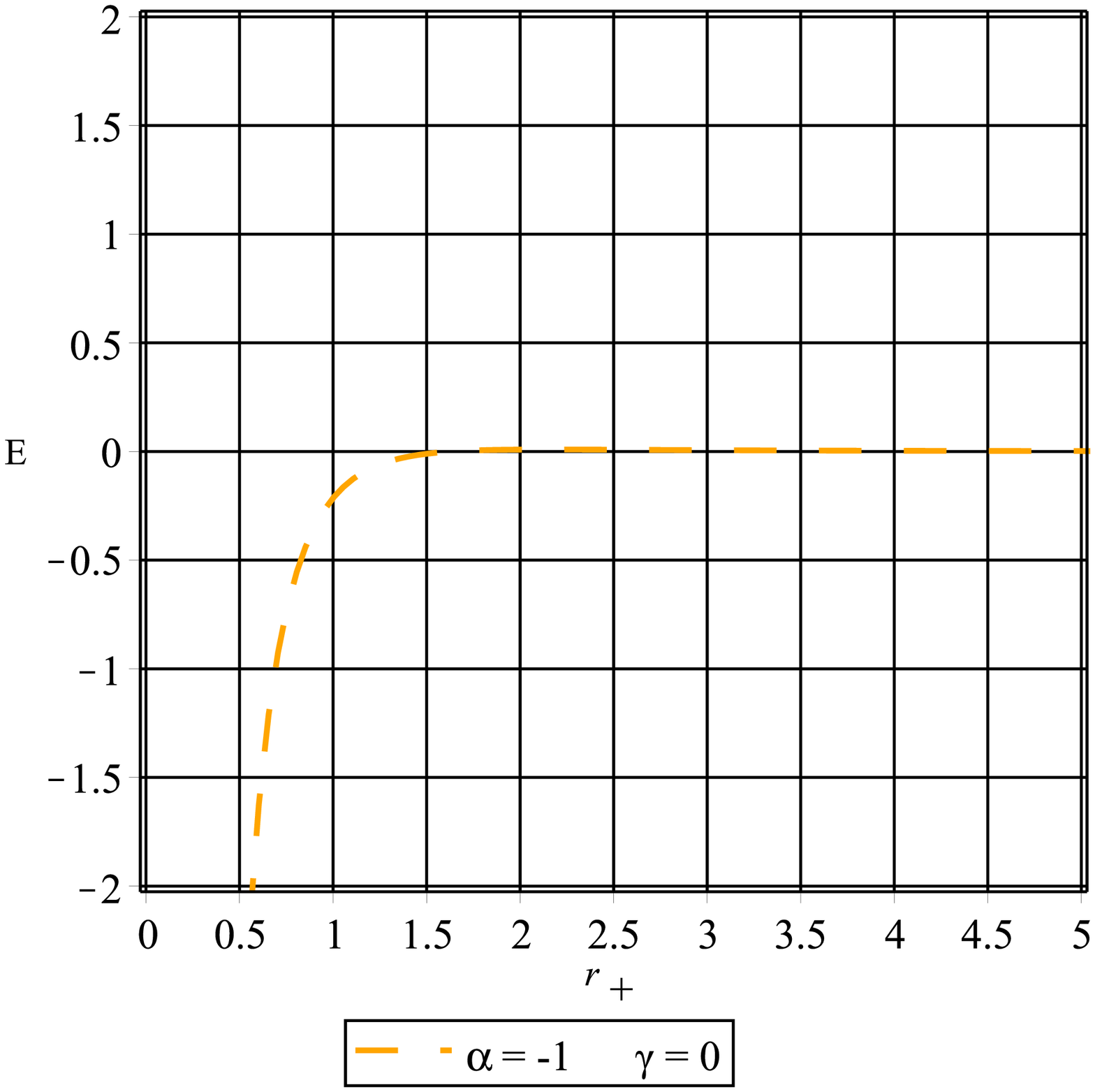}\includegraphics[scale=.3]{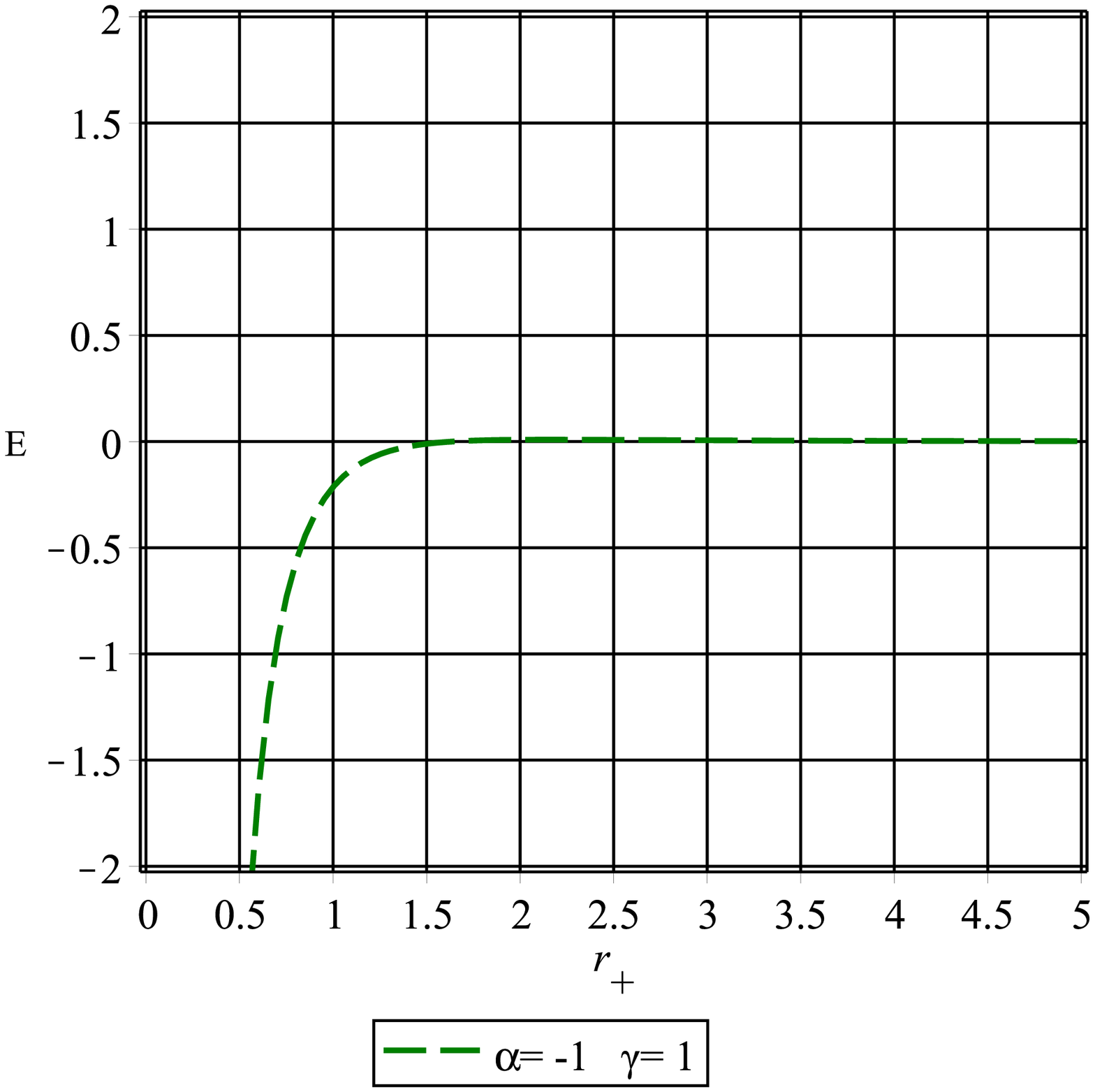}\\
\includegraphics[scale=.3]{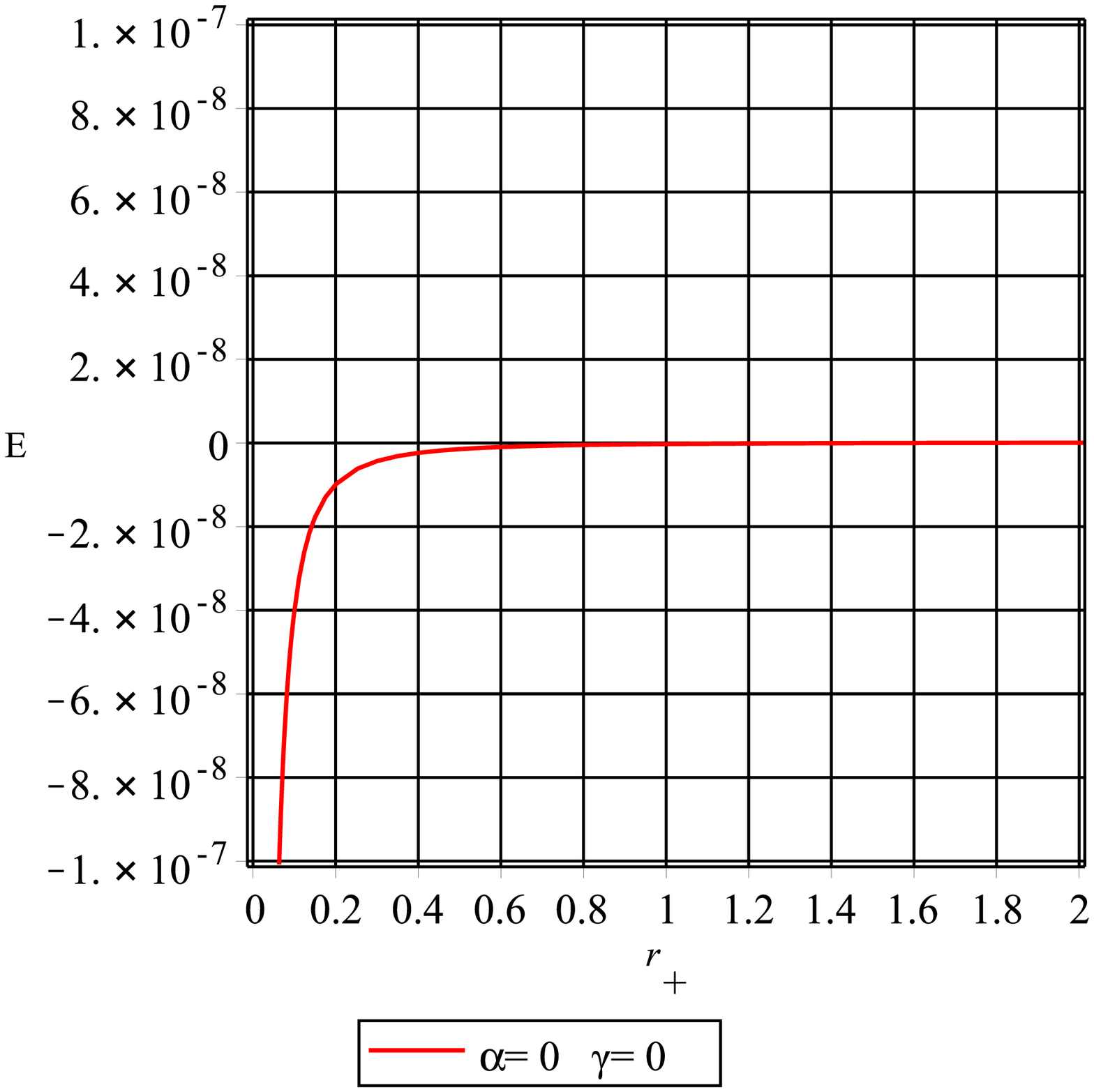}\includegraphics[scale=.3]{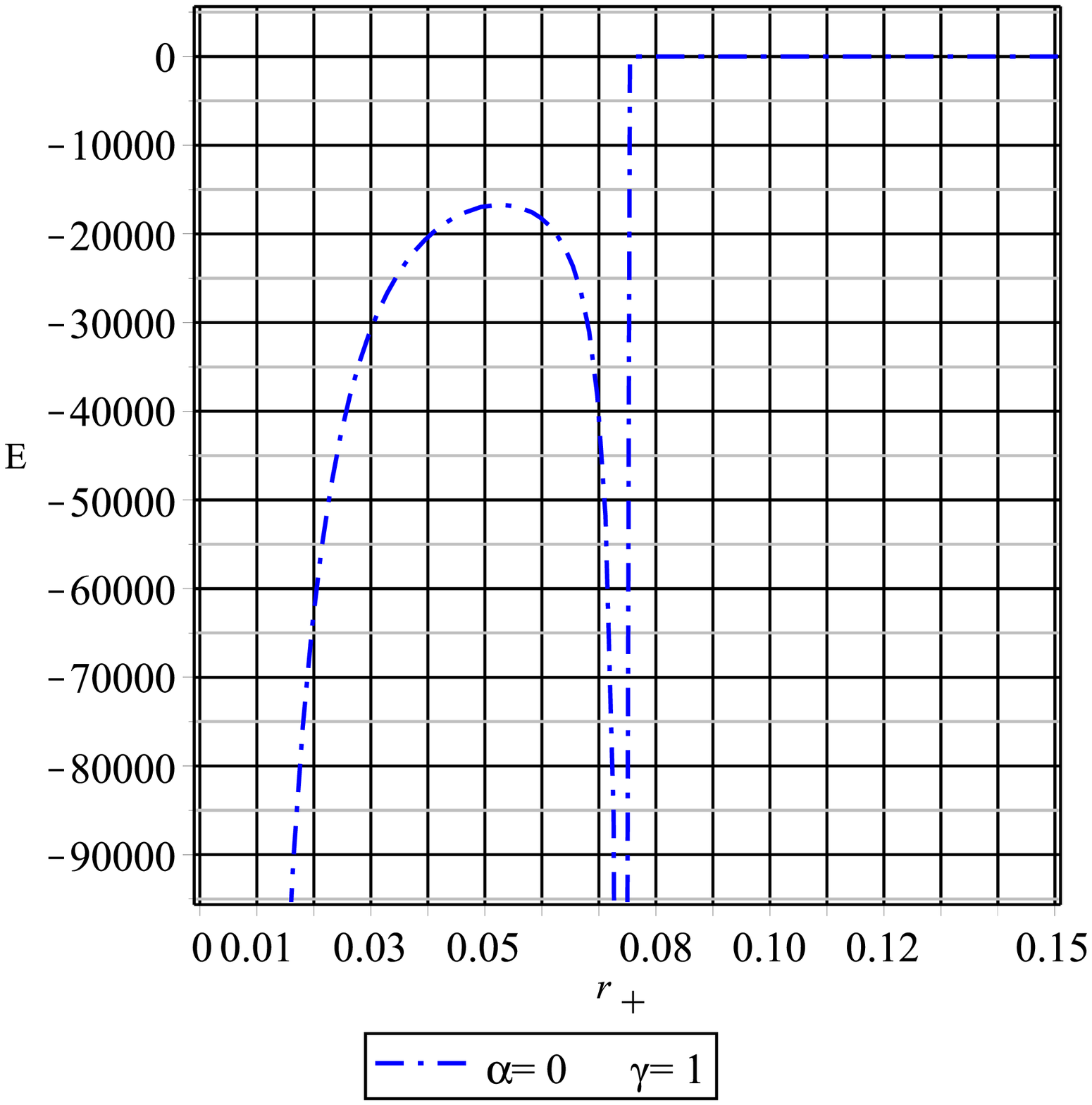}\\
\caption{Internal energy according to the radius of the event horizon for values $\mu=1$, $q=0.1$, $a=1$ and $j=0.0649$.}
\label{fig2}
\end{center}
\end{figure}

We draw the internal energy in the presence and absence of correction terms. When we have the first sentence of the principles of entropy, the presence and absence of the second sentence does not have any effect on the internal energy of the system, but when we ignore the first of these reforms, the presence and absence of the second sentence of the reform will make a lot of changes in energy. This effect can be seen in the following form. In the third plot of the Fig. \ref{fig2}, both terms are discarded, and the entropy is equal to the original one. But in the last plot, considering the second sentence, energy changes range from -100,000 to -15,000. The maximum value of these changes is -15000 for $r_{+} = 0.05$. As we approach $r_{+} = 0.08$, the energy decreases as an asymptotic line, so that $r_{+} = 0.08$ can be considered as an asymptotic of the graph.

\subsection{Helmholtz free energy}
\begin{figure}[th]
\begin{center}
\includegraphics[scale=.5]{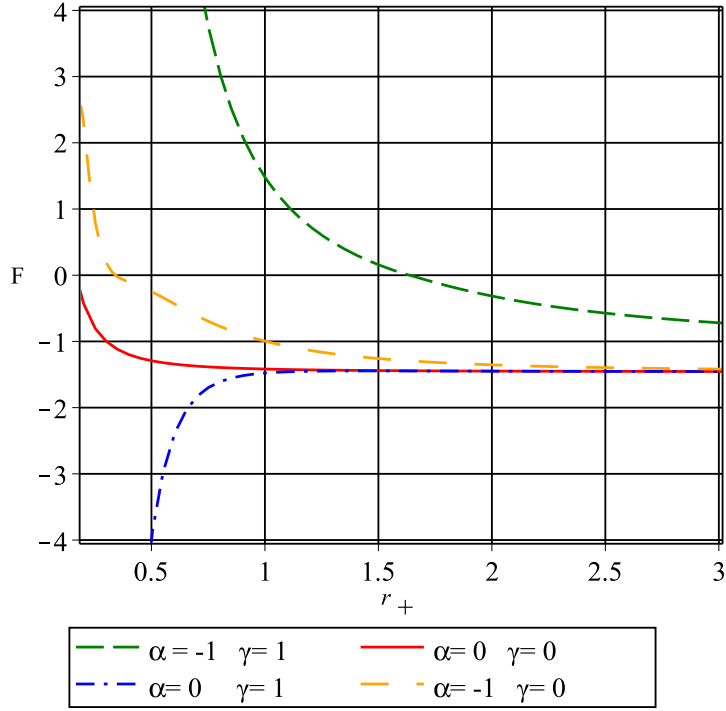}
\caption{Helmholtz free energy according to the radius of the event horizon for values $\mu=1$, $q=0.1$, $a=1$, and $j=0.0649$.}
\label{fig6}
\end{center}
\end{figure}

Helmholtz free energy can also be obtained using the entropy and the partition function as follow,
\begin{equation}\label{19}
{F}=-{k}_{B}{T}{\ln Z},
\end{equation}
so we can find its changes in the presence and absence of correction terms as illustrated by the Fig. \ref{fig6}.

Fig. \ref{fig6} shows the free energy of the black hole regarding the radius of the event horizon. Whatever $r_{+}=0$ go to $r_{+}=3$, Helmholtz free energy goes from $F=0$ to $F\simeq-1\cdot5$, which can be called a curved asymmetric line.\\
When we add the second term of the Helmholtz free energy, energy starts at very low values and loses its effect at $r_{+}\leq1$. Therefore, it can be concluded that the second term only has an effect on Helmholtz free energy at $0\leq r_{+} \leq 1\cdot5$.\\
Now, if we only consider the effect of the first term of corrections of the entropy, free energy starts from $F\simeq3$, and its effect on the entropy is up to $r_{+}\leq2\cdot5$. The point is that this effect will increase Helmholtz free energy, which is in the contrary with the effect of the second term.\\
Now, if we consider both corrective sentences, we see significant changes in the graph. First, there is no starting point in the chart. This means that energy is infinite at the starting point. Perhaps it can be analyzed that at $r_{+}=0$, Helmholtz free energy is meaningless, just like the energy at zero point in harmonic oscillators. Secondly, at $r_{+}\simeq1\cdot6$ we can see vanishing of the Helmholtz free energy. Thirdly, in $1\cdot6\leq r_{+} \leq 3$ the graph goes toward unmodified Helmholtz free energy. But the obvious difference between this graph with the one we did not consider the corrections is that in the first three modes at $r_{+}=3$ all the graphs represent an energy; However, we consider both corrective sentences, Helmholtz free energy is about one unit higher than the unmodified state, which is a positive change indicating that the system is reversible.

\section{The first law of the black hole thermodynamics}
\begin{figure}[th]
\begin{center}
\includegraphics[scale=.25]{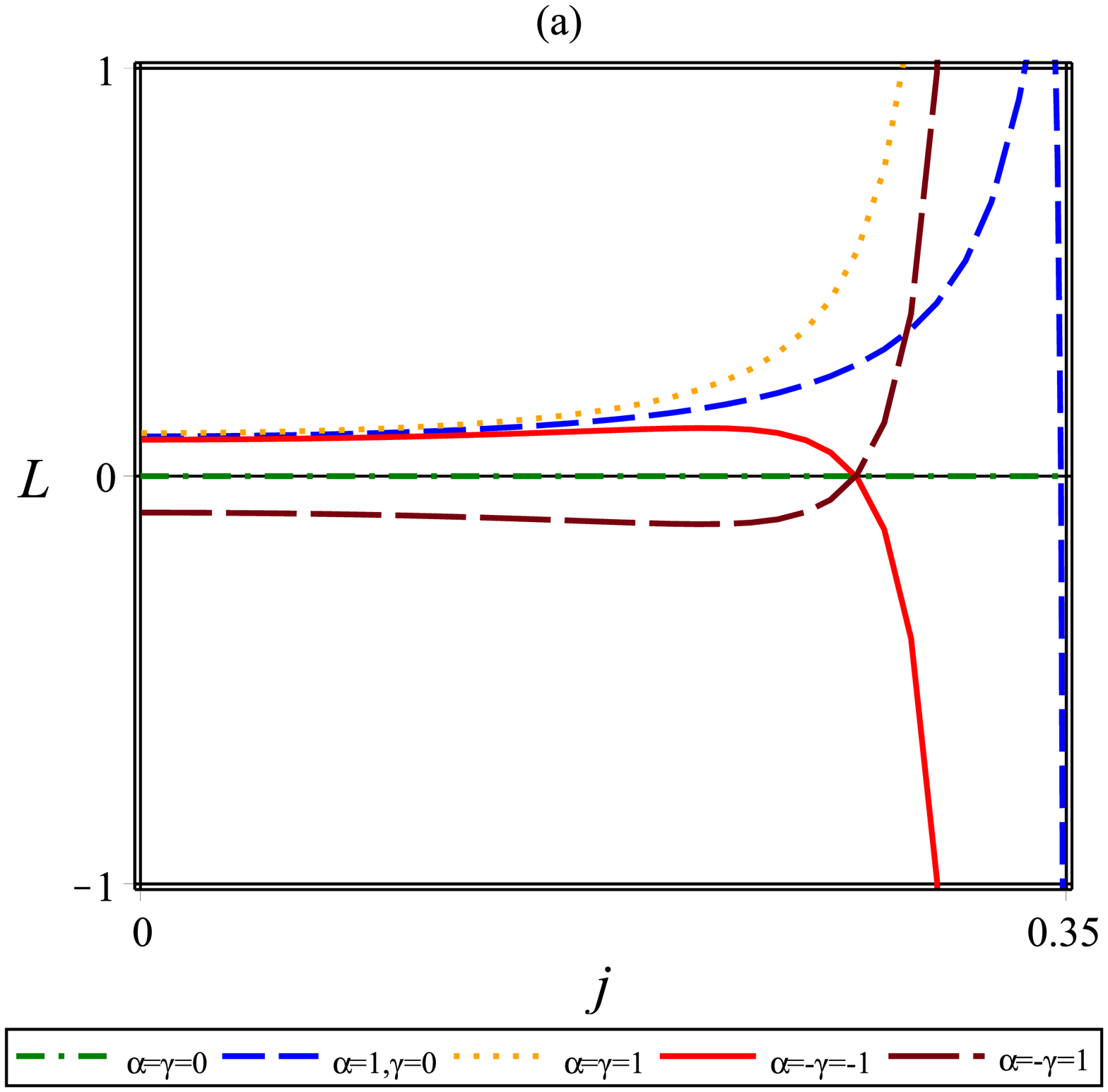}\includegraphics[scale=.25]{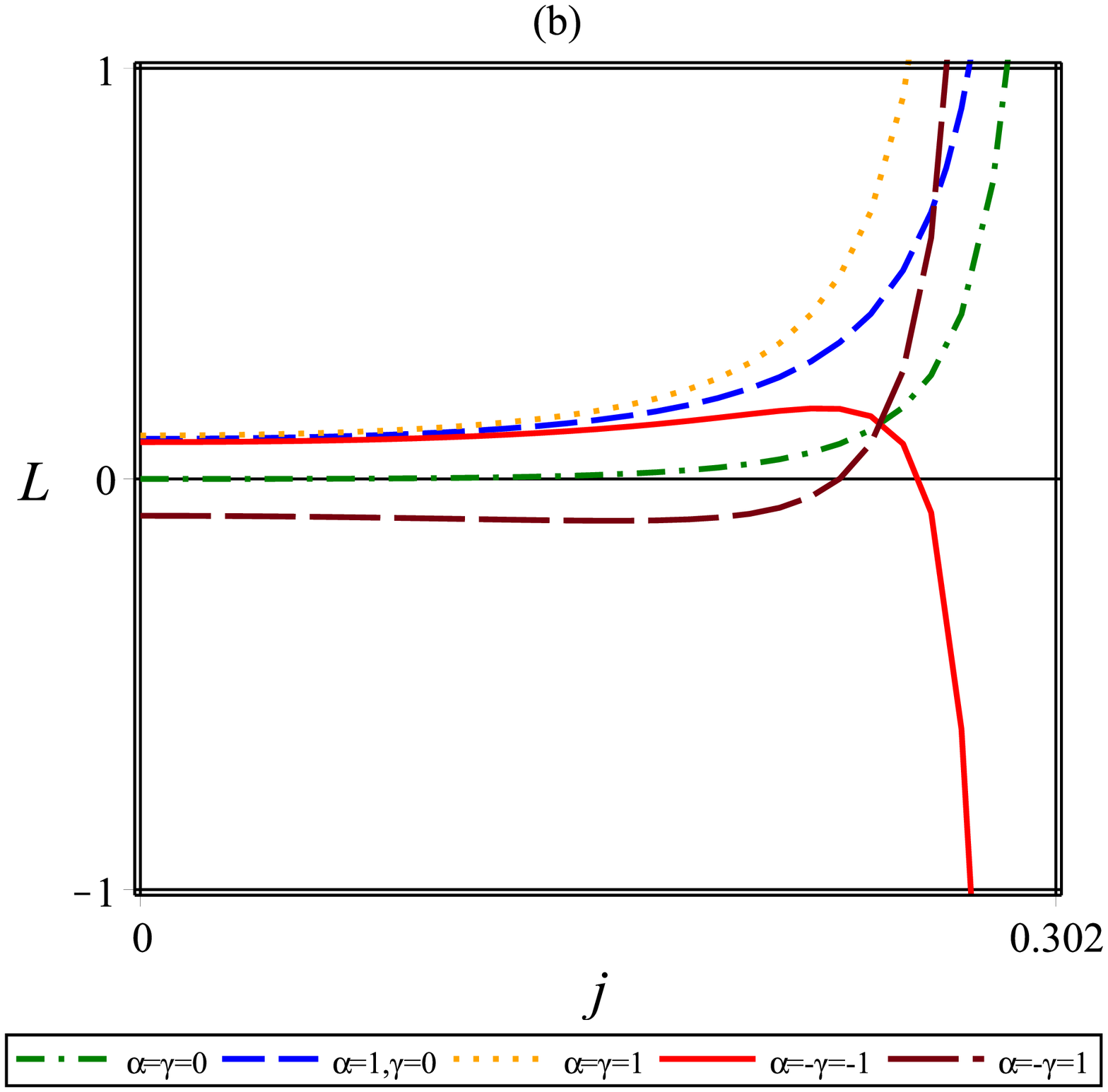}\includegraphics[scale=.25]{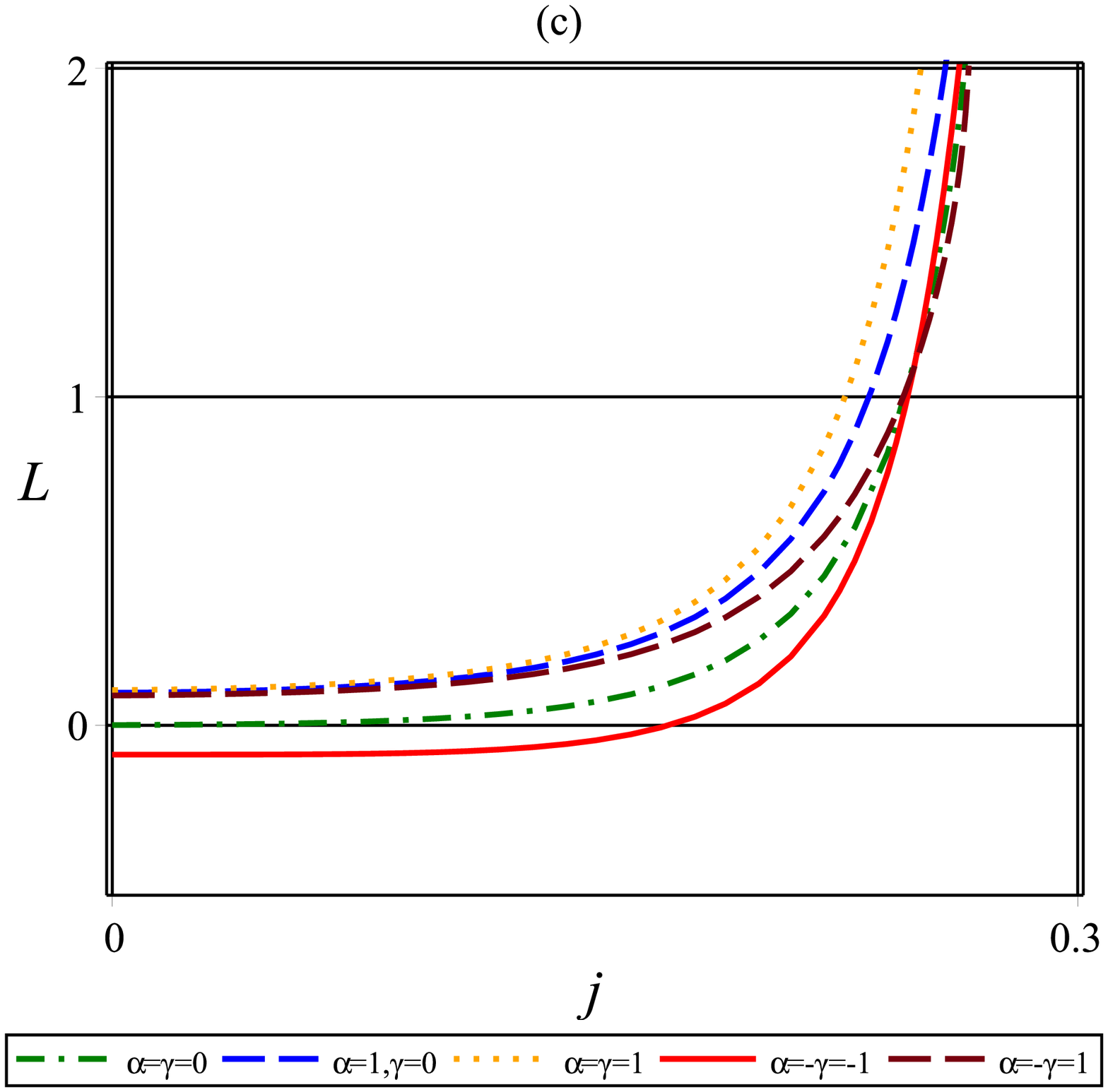}
\caption{Plots of $L$ in terms of $j$ for values $\mu=1$, and (a) $a=q=0$; (b) $q=0.1$ and $a=0$; (c) $a=q=0.1$.}
\label{figL}
\end{center}
\end{figure}

In the Refs. \cite{014, 1101} it is argued that the thermodynamics quantities of the charged rotating G\"{o}del black hole satisfy the first law of thermodynamics, hence the following relations hold for $\alpha=\gamma=0$,
\begin{equation}\label{4-1}
dM=TdS+\Omega dJ+\Phi dQ+Wdj
\end{equation}
and
\begin{equation}\label{4-2}
\frac{2}{3}M=TS+\Omega J+\frac{2}{3}\Phi Q-\frac{1}{3}Wj,
\end{equation}
where the mass calculated as \cite{014},
\begin{equation}\label{4-3}
M=\pi\left(\frac{3}{4}\mu-j(\mu+q)a-2j^{2}(m+q)(4m+5q)\right),
\end{equation}
while angular momenta given by \cite{014},
\begin{equation}\label{4-4}
J=\frac{\pi}{2}\left[a\left(\mu-\frac{q}{2}-2j(\mu-q)a-8j^{2}(\mu^{2}+\mu q-2q^{2})\right)-3jq^2+8j^{2}(3\mu+5q)q^2\right],
\end{equation}
with its conjugate variable given by the equation (\ref{5}) which calculated at $r=r_{+}$. Moreover, the conserved charge obtained as \cite{014},
\begin{equation}\label{4-5}
Q=\frac{\sqrt{3}\pi}{2}\left[q-4j(\mu+q)a-8j^2(m+q)q\right],
\end{equation}
which has the following electrostatic potential as its conjugate,
\begin{equation}\label{4-6}
\Phi=\frac{\sqrt{3}}{2}\left(\frac{q}{r_{+}^{2}}+(jr_{+}^{2}+2jq-\frac{qa}{2r_{+}^{2}})\Omega(r_{+})\right).
\end{equation}
Finally, the generalized force
\begin{equation}\label{4-7}
W=2\pi(\mu+q)\left(a+2j(\mu+2q)\right),
\end{equation}
is conjugate variable of the G\"{o}del parameter.\\
Now we would like to investigate relation (\ref{4-2}) for the above quantities together the temperature (\ref{8}) and corrected entropy (\ref{15}).
In order to give numerical analysis of the equation (\ref{4-2}) we rewrite it as follow,
\begin{equation}\label{4-2}
L=TS+\Omega J+\frac{2}{3}\Phi Q-\frac{1}{3}Wj-\frac{2}{3}M=0,
\end{equation}
and plot $L$ in terms of some parameters in the Fig. \ref{figL} for three different cases of G\"{o}del, charged G\"{o}del, and charged rotating  G\"{o}del black holes.

We can see that the first law of the black hole thermodynamics hold ($L=0$) for the case of $\alpha=\gamma=0$ (dash dotted (green) line of plots). Plot (a) shows ordinary G\"{o}del black hole (uncharged static), plot (b) shows charged G\"{o}del black hole without rotation and plot (c) shows charged rotating G\"{o}del black hole. In all cases we can see that the first law of black hole in presence of the logarithmic corrected entropy (first order correction with $\gamma=0$) violated (see dashed blue line). In the case of higher order corrected entropy we can see there is special choice for the $j$ parameter which yields to $L=0$ hence satisfy the first law of thermodynamics (for example see solid red line of plots).

\section{Heat capacity}
One of the best ways to investigate the stability of black hole is the study of the sign of specific heat, while it is also possible to study the stability of black holes based on horizon thermodynamics \cite{PLB3}. The heat capacity is written using the black hole entropy and temperature as follow,
\begin{equation}\label{25}
{C}={T}{{d}{S} \over {d}{T}},
\end{equation}
which we examine for the black hole stability in the presence of logarithmic and higher order entropy corrections.

\begin{figure}[th]
\begin{center}
\includegraphics[scale=.5]{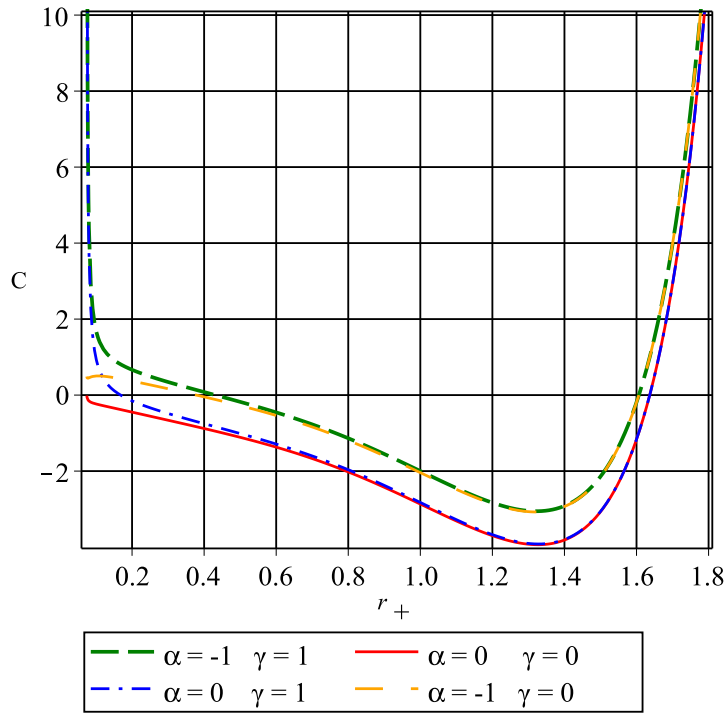}
\caption{Heat capacity according to the radius of the event horizon for values $\mu=1$, $q=0.1$, $a=1$, and $j=0.0649$.}
\label{fig10}
\end{center}
\end{figure}

Fig. \ref{fig10} shows the heat capacity diagram in terms of the radius of the event horizon. In the absence of corrective  sentences, the black hole is unstable $(r_{+}<0\cdot65)$, but with the first term, the graph is slightly upward, so that for $r_{+}<0\cdot4$ the amount of heat capacity is positive. Considering the second sentence, the upsurge for $r_{+}<0\cdot2$ is too high. So, in general, it can be concluded that by modifying the black hole from unstable to steady state. In this diagram, the amount of heat capacity is not considered, but it is merely a sign of controversy, as illustrated by the Fig. \ref{fig10}. For $r_{+}>1\cdot65$, the heat capacity is positive, which means the stability of the Kerr-Newman-G\"{o}del black hole. We can obtain similar result by taking $j$ as variable parameter and use the equation (\ref{7}) for the event horizon radius.

\section{Statistical mechanics}
By using the thermodynamic quantities, we can study the black hole statistics.

\subsection{Micro-states}
The thermodynamics and statistical mechanics are linked by the following equation
\begin{equation}\label{26}
{S}={k}{\ln}{\Omega},
\end{equation}
where $\Omega$ is the number of micro-states of the system.

\begin{figure}[th]
\begin{center}
\includegraphics[scale=.3]{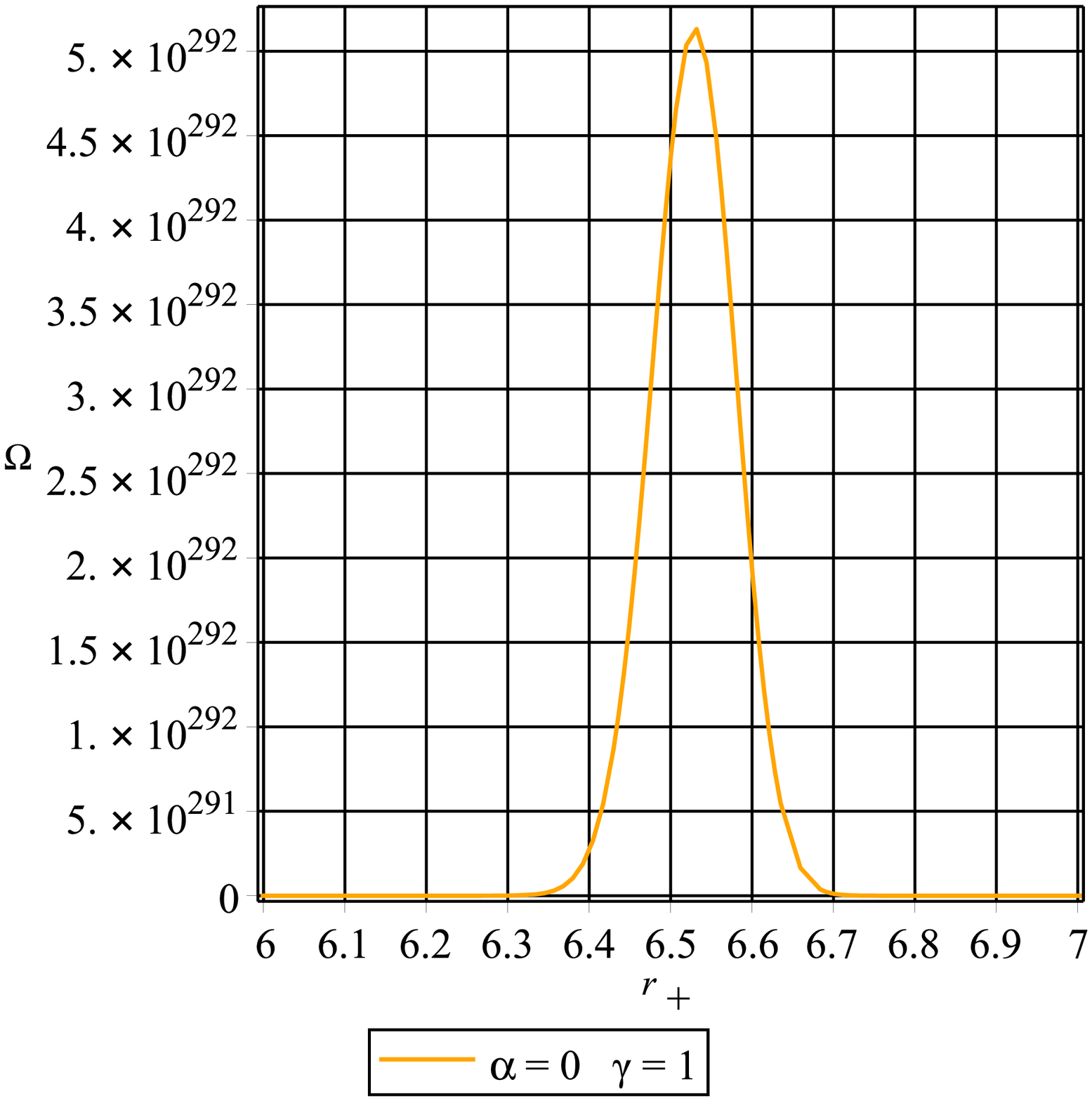}\includegraphics[scale=.3]{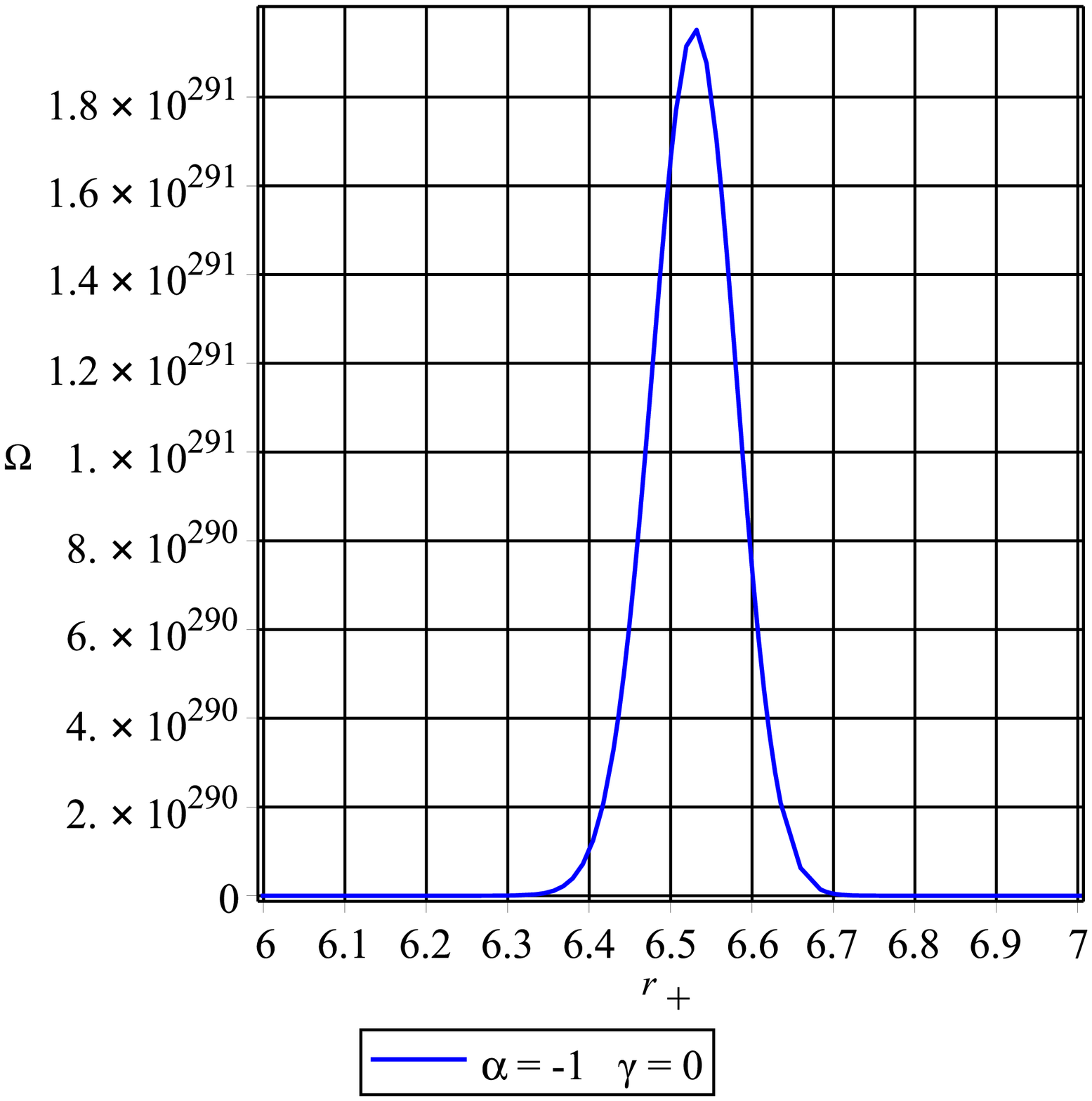}\\
\includegraphics[scale=.3]{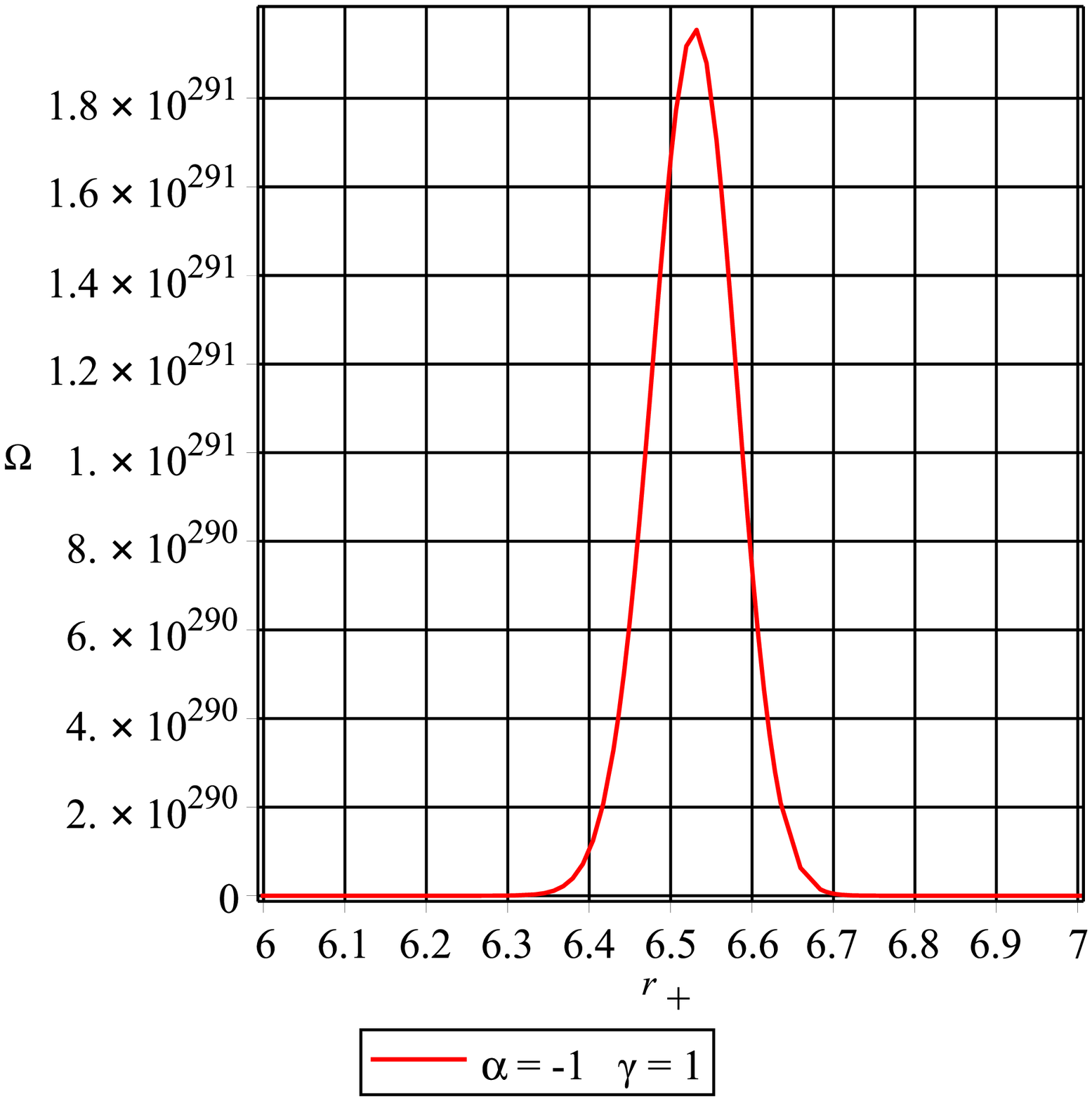}\includegraphics[scale=.3]{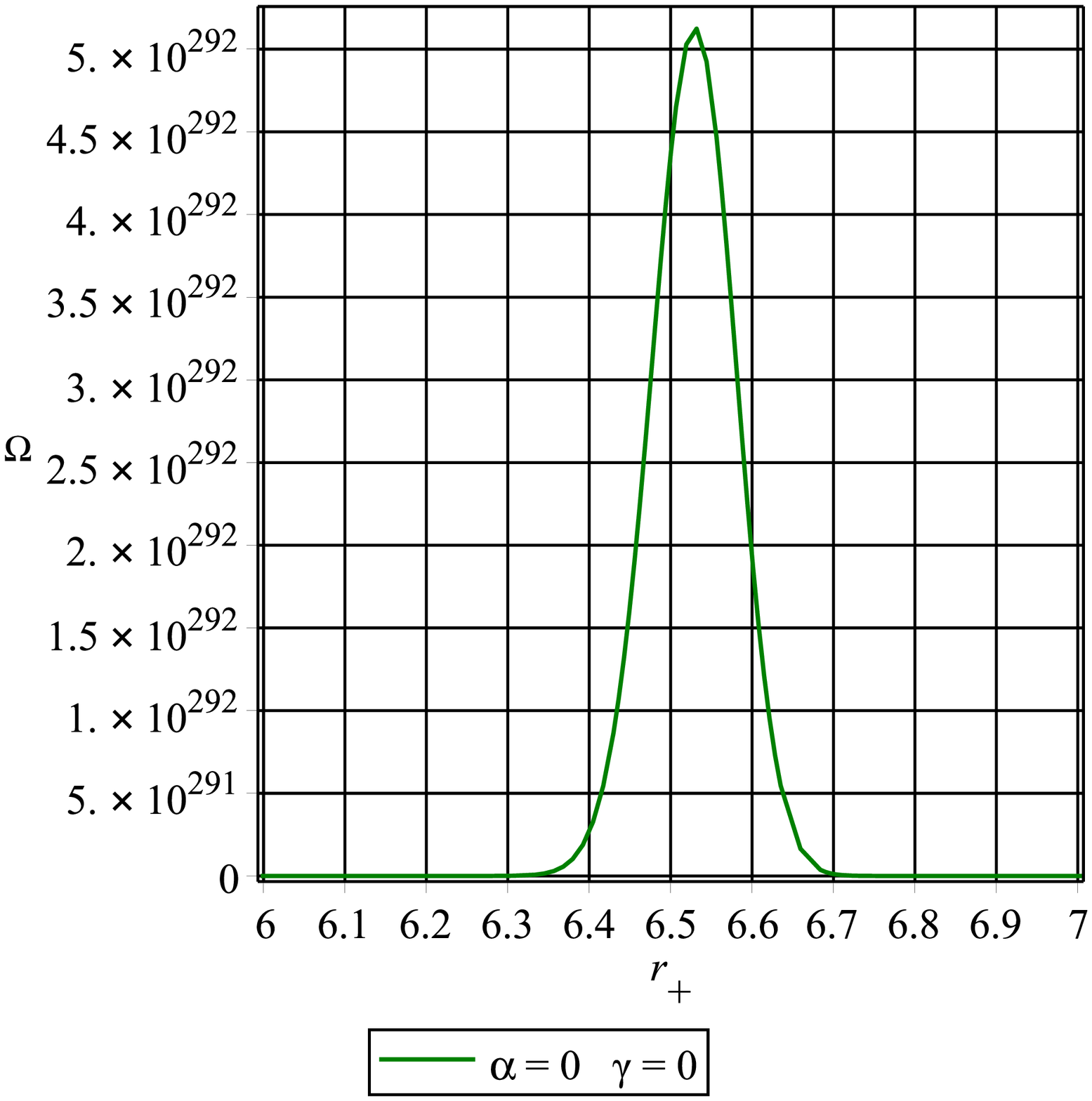}
\caption{Micro-states according to the radius of the event horizon for values $M=1$, $q=0.1$, $a=1$, and $j=0.0649$.}
\label{fig11}
\end{center}
\end{figure}

Plots of the Fig. \ref{fig11} show the micro-states diagram of the system in terms of the radius of the event horizon. As seen in the figure, in the absence of any corrected expressions, the peak diagram for $r_{+}\simeq6\cdot53$ equals $5\times{10}^{292}$ (third plot).\\
The system in the range $6\cdot3<r_{+}<6\cdot7$ has a micro-states. In the presence of only the second correction term, and ignoring the first sentence, the micro-states diagram does not change, that is, the second correction sentence does not affect the micro-states of the system (second plot).\\
In the presence of only the first term, the correction of the peak diagram in the point is reduced to $27\cdot7$ times (first plot), Therefore, considering both corrections (last plot) and the fact that only the first term in the diagram affects, the diagram is when there are both terms, flows toward a diagram that is only included in the first term. It can be concluded that entropy corrections reduce $27\cdot7$ times the micro-states of the system.

\subsection{Partition function}
By using the relation (\ref{17}) one can obtain the black hole partition function and study effects of higher order corrections of entropy.

\begin{figure}[th]
\begin{center}
\includegraphics[scale=.3]{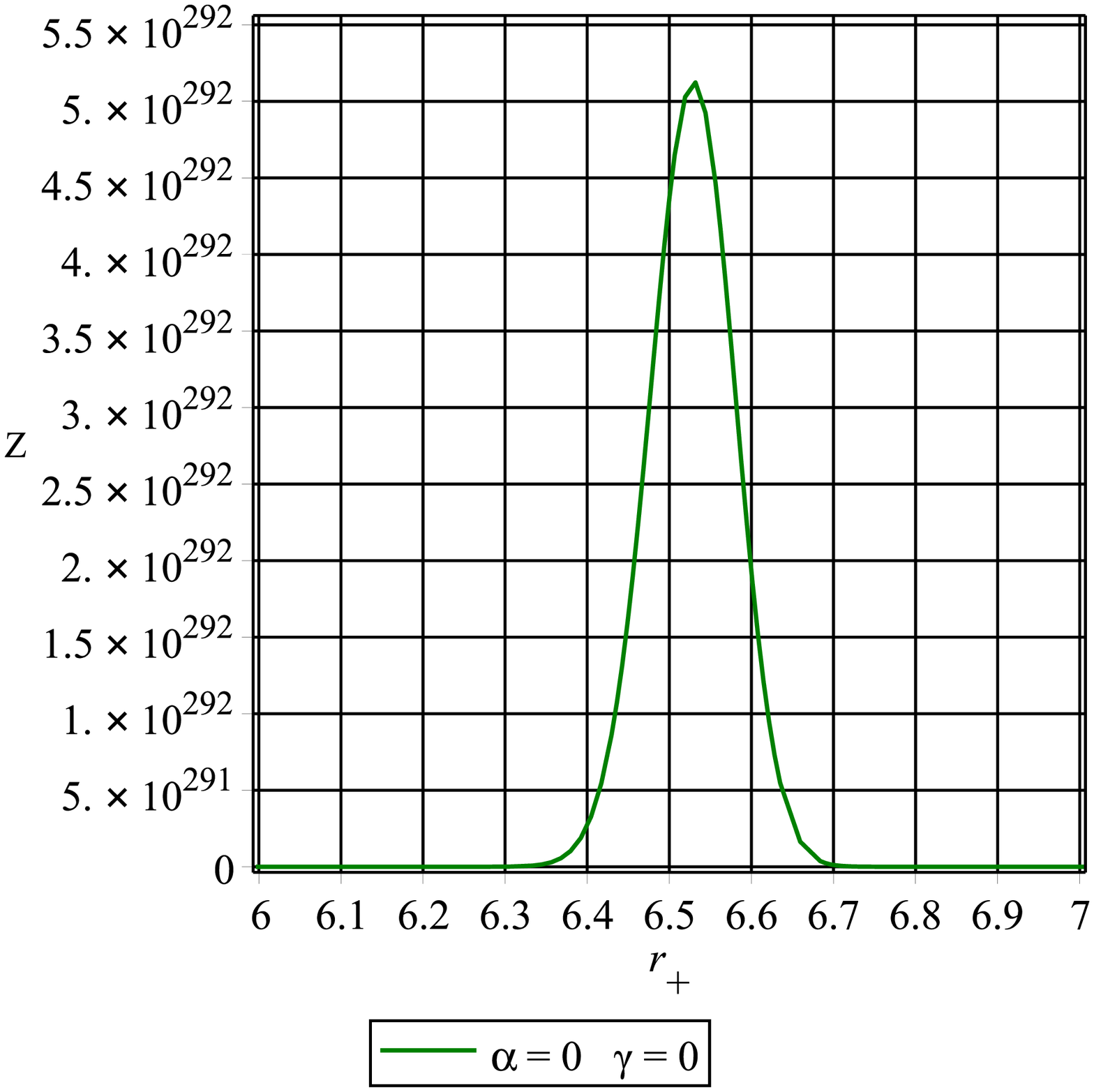}\includegraphics[scale=.3]{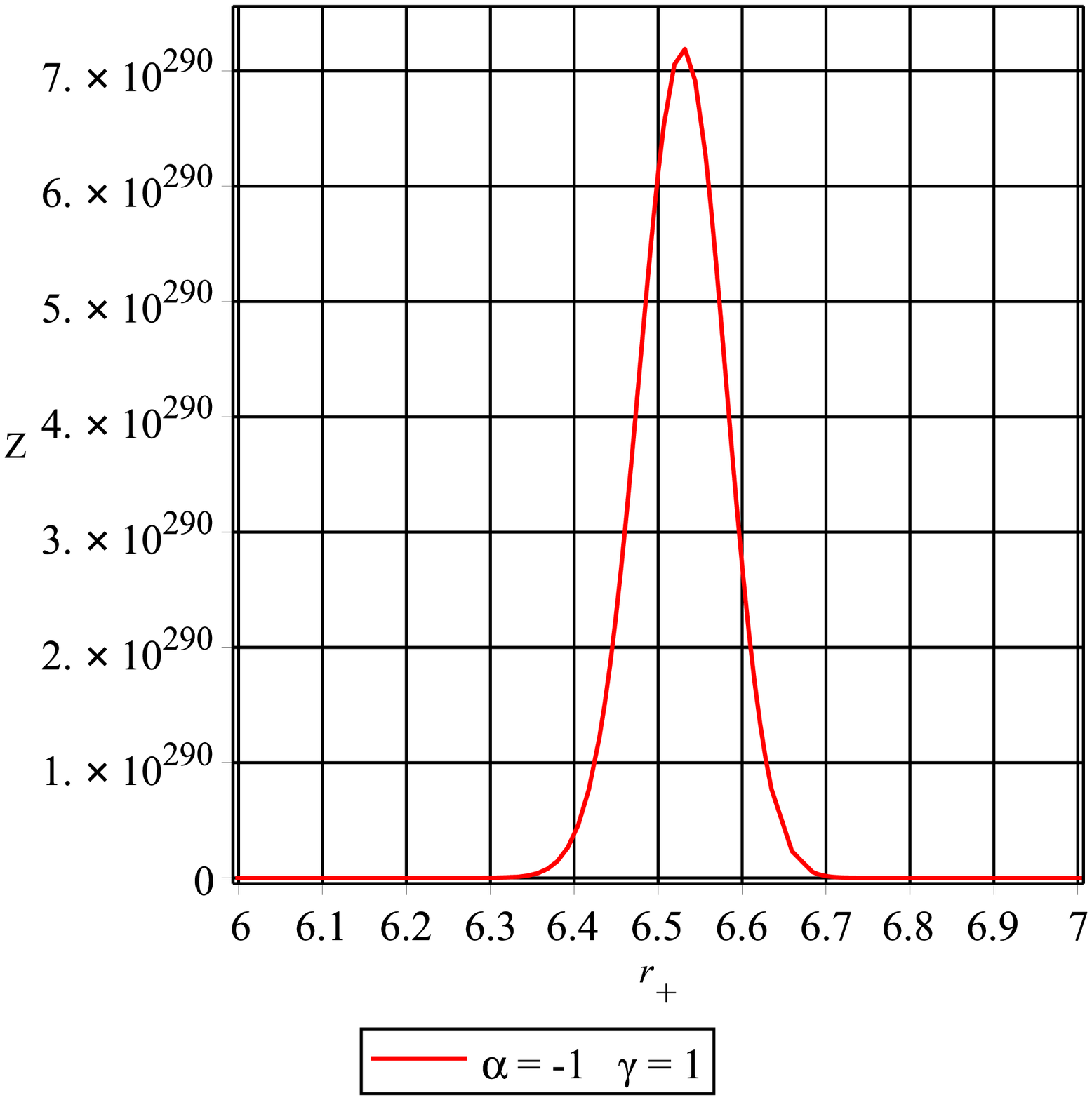}\\
\includegraphics[scale=.3]{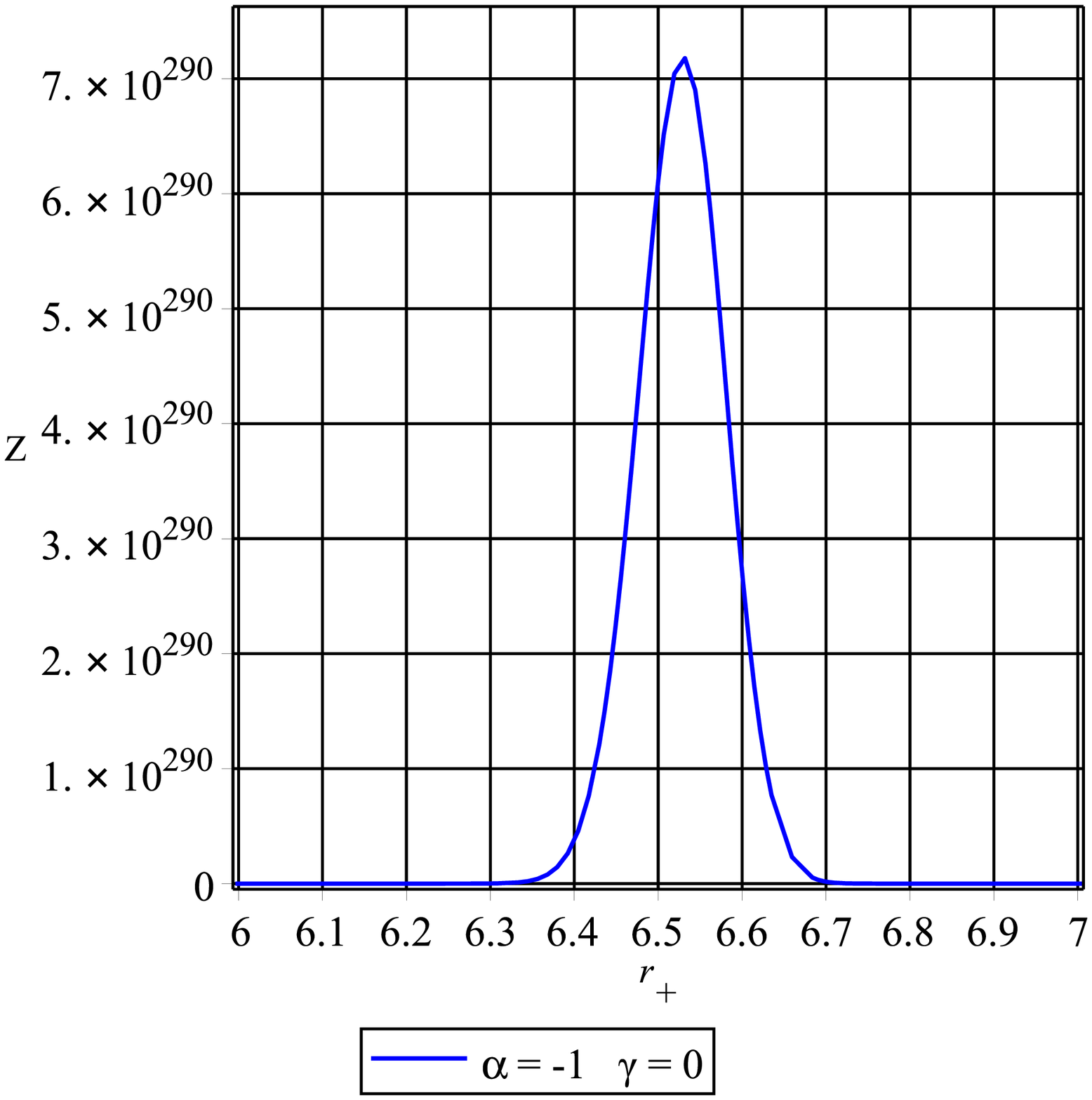}\includegraphics[scale=.3]{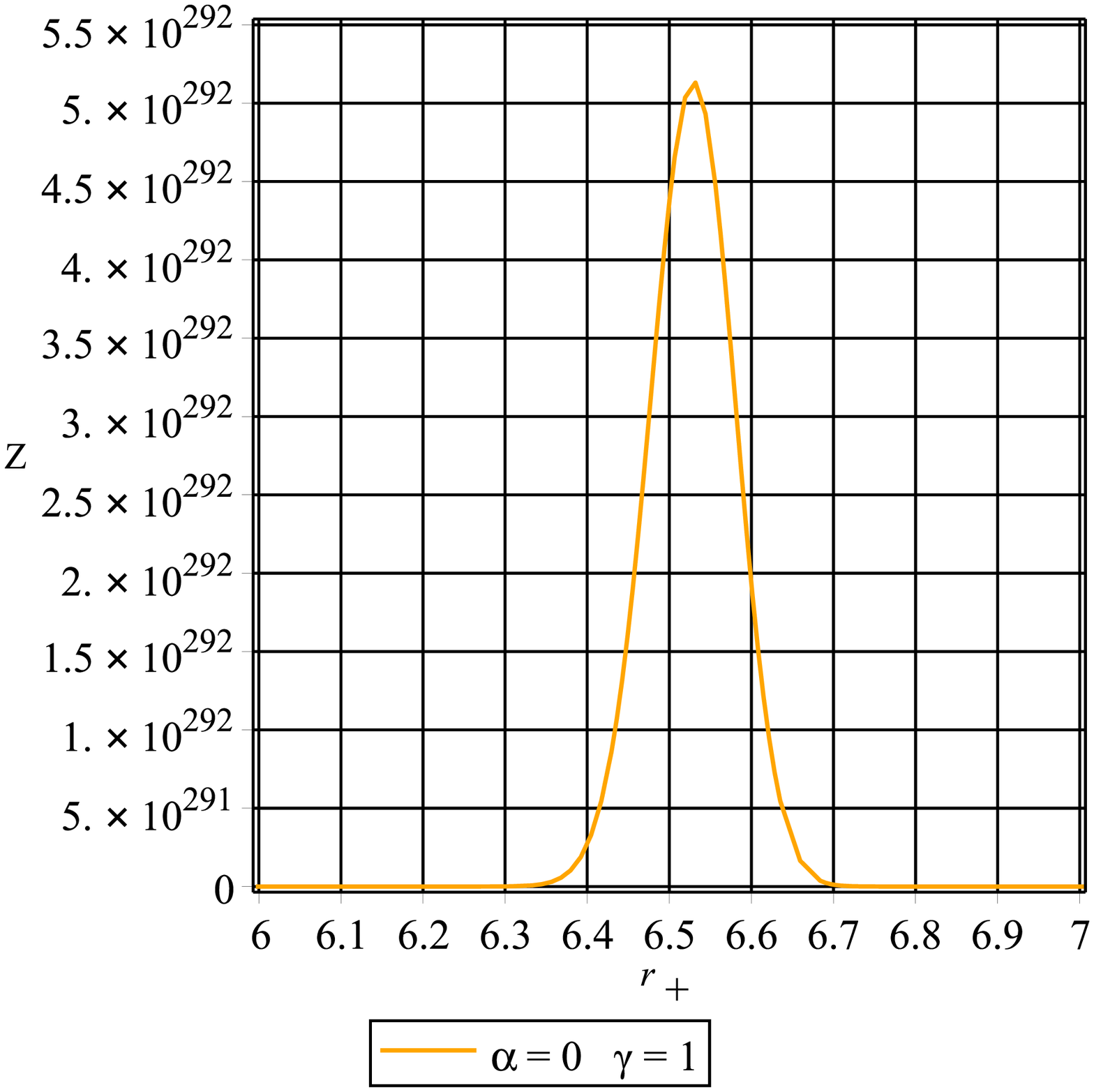}
\caption{Partition function according to the radius of the event horizon for values $\mu=1$, $q=0.1$, $a=1$, and $j=0.0649$.}
\label{fig15}
\end{center}
\end{figure}

Plots of the Fig. \ref{fig15} show the partition function of the system in terms of the radius of the event horizon. The partition function gives the statistical characteristics of a system in the thermodynamic equilibrium. This function is a function of temperature and volume, And given that every point of a black hole has a certain temperature and volume, it can be attributed to each point of the black hole a specific partition function.\\
Regardless of correction sentences, the peak diagram is located at $5\times{10}^{292}$, This point is the same as the micro-states courier's place.\\
When we only consider the second term, diagram partition function does not change. In other words, the second corrective sentence has no effect on the system function.\\
In the presence of the first correction term, the peak diagram in the point is reduced.\\
So, considering both correction terms and given that only the first term affects the diagram of the partition function, and graph in the presence of both correction terms goes to the graph that only affected the first term. 

\subsection{Probability function}
Using the following equation, one can obtain the probability that the system will be in an energy state ${E}_{r_{+}}$,
\begin{equation}\label{27}
{P}={{\exp}{(-\beta{E}_{r})} \over {{\sum}_{r}{\exp}{(-{\beta}{{E}_{r}})}}}={{\exp}{(-{\beta}{{E}_{r}})} \over {Z}}.
\end{equation}

\begin{figure}[th]
\begin{center}
\includegraphics[scale=.5]{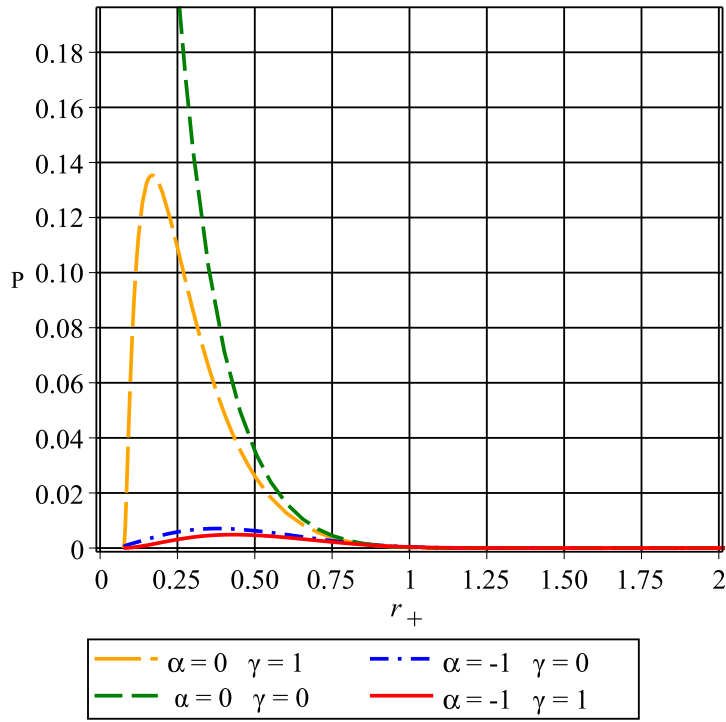}
\caption{Probability function according to the radius of the event horizon for values $\mu=1$, $q=0.1$, $a=1$, and $j=0.0649$.}
\label{fig19}
\end{center}
\end{figure}

Fig. \ref{fig19} shows the probability function in terms of the radius of the event horizon. In the absence of correction sentences for $r_{+}<0.25$, the probability function has a large amount, and this value goes up to $100$, but never reaches this number; The reason is that we never have a black hole in zero radius.\\
As $r_{+}=0.25$ goes to $r_{+}=0.75$, the probability function comes from $P\simeq0.2$ to $P\simeq0$ and for a radius larger than one, the probability function is zero, This means that the other system is not in the energy state ${E}_{r_{+}}$.\\
By applying the second-order correction, the probability function decreases slightly so that its maximum is in $r\simeq0.2$. If we only consider the first term, the probability function falls sharply and does not exceed $0.01$ in all radius.\\
Finally, taking into account all the corrections, the maximum probability function is less than $0.01$, And from a radius of roughly $r_{+}>1$, the probability function tends to zero; this means that the probability that the system will be in a state with ${E}_{r_{+}}$ is almost zero, That is, the system energy must be infinite.

\section{Conclusions}
In this paper, we examined the effect of entropy corrections on thermodynamics and statistical mechanics of Kerr-Newman-G\"{o}del black hole, and observed that in the presence of both correction terms, the entropy of the system is completely different, and the entropy of the system is greatly increased. The leading order correction is logarithmic \cite{3}, while higher order corrections are proportional to the inverse of the entropy.
The internal energy of the system is significant in the presence of the first single sentence, and this has a significant role in the internal energy.
We considered both correction terms and found the Helmholtz free energy is about one unit higher than the unmodified state, which is a positive change indicating that the system is reversible.
We found that the first law of the black hole thermodynamics may hold with higher order corrections while violate with the logarithmic (first order) correction.
In the radial range, the second term does not affect the micro-states and partition of the system, but the first term (logarithmic term) has a significant effect, the micro-states and partition of the system are greatly reduced.
In the presence of both correction terms, the probability function is greatly reduced than when we did not enter them.
In a small radius, the heat capacity, when we do not enter correction term, is completely negative in small radius, but when we enter both terms, the heat capacity is positive and the black hole goes to the stable phase. It tells that correction terms are many important to have stable Kerr-Newman-G\"{o}del black hole. In that case it would be interesting to investigate Smarr formula \cite{CQG} for the Kerr-Newman-G\"{o}del black hole in presence of higher order corrections of the entropy.

\end{document}